\documentclass[12pt]{article}
\usepackage{graphicx}
\usepackage[cp1251]{inputenc}

\begin{document}
 \noindent {\footnotesize\it Astronomy Letters, 2019, Vol. 45, No 4, pp. 208--216.}
 \newcommand{\dif}{\textrm{d}}

 \noindent
 \begin{tabular}{llllllllllllllllllllllllllllllllllllllllllllll}
 & & & & & & & & & & & & & & & & & & & & & & & & & & & & & & & & & & & & & \\\hline\hline
 \end{tabular}

  \vskip 0.5cm
  \centerline{\bf\large Kinematic Properties of Open Star Clusters with Data}
  \centerline{\bf\large from the Gaia DR2 Catalogue}
   \bigskip
  \bigskip
  \centerline
 {V.V. Bobylev\footnote [1]{e-mail: vbobylev@gaoran.ru} and A.T. Bajkova}
  \bigskip

  \centerline{\small\it Pulkovo Astronomical Observatory, Russian Academy of Sciences,}

  \centerline{\small\it Pulkovskoe sh. 65, St. Petersburg, 196140 Russia}
 \bigskip
 \bigskip
 \bigskip

 {
{\bf Abstract}---We consider open star clusters (OSCs) with the
proper motions, parallaxes, and line-of-sight velocities
calculated by various authors from Gaia DR2 data. The distance
scale factor has been found by analyzing the separate solutions of
the basic kinematic equations to be $p=1.00\pm0.04.$ It shows that
the distances calculated using the parallaxes from the Gaia DR2
catalogue do not need any correction factor. We have investigated
the solutions obtained from various OSC samples differing in both
age and accuracy of their parallax and line-of-sight velocity
measurements. The solution obtained from a sample of 930 OSCs
satisfying the constraints on the age $\log t<9$ and the relative
trigonometric parallax error $<$30\% is recognized to be the best
one, with 384 OSCs in this sample having the mean line-of-sight
velocities calculated from at least three probable members of the
corresponding clusters. As a result of the simultaneous solution
of all the basic kinematic equations, we have found the following
kinematic parameters from this sample:
 $(U,V,W)_\odot=(8.53,11.22,7.83)\pm(0.38,0.46,0.32)$ km s$^{-1}$,
 $\Omega_0=28.71\pm0.22$ km s$^{-1}$ kpc$^{-1}$,
 $\Omega'_0=-4.100\pm0.058$ km s$^{-1}$ kpc$^{-2}$, and
 $\Omega''_0= 0.736\pm0.033$ km s$^{-1}$ kpc$^{-3}$.
 The linear rotation velocity at the adopted at the adopted solar distance
 $R_0=8\pm0.15$~kpc is
 $V_0=229.7\pm4.6$ km s$^{-1}$.
 An analysis of the proper motions for these 930 OSCs has
shown that, apart from the rotation around the Galactic z axis,
there is rotation of the entire sample around the x axis with an
angular velocity of $0.48\pm0.15$ km s$^{-1}$ kpc$^{-1}$ differing
significantly from zero.
  }


 \subsection*{INTRODUCTION}
Open star clusters (OSCs) play an important role in studying the
Galaxy and its subsystems, because they have a high accuracy of
the mean values of a number of kinematic and photometric
parameters. The accuracy of the distances, proper motions, and
line-of-sight velocities of OSCs is of great importance here.

The Gaia DR2 catalogue containing the trigonometric parallaxes and
proper motions of $\sim$1.3 billion stars was published in 2018
(Brown et al. 2018; Lindegren et al. 2018). The mean parallax
errors lie in the range 0.02--0.04 milliarcseconds (mas) for
bright $(G<15^m)$ stars and reach 0.7 mas for faint $(G=20^m)$
stars. For more than 7 million stars of spectral types F--G--K the
line-of-sight velocities were measured with a mean error of
$\sim$1 km s$^{-1}$.

Using these up-to-date data is of great importance for studying
the kinematics of stars and clusters. Based on them, the
structural and kinematic parameters of a large number of OSCs have
been refined (Babusiaux et al. 2018; Kuhn et al. 2018;
Cantat-Gaudin et al. 2018), the properties of a number of young
stellar associations (Zari et al. 2018; Franciosini et al. 2018;
Roccatagliata et al. 2018; Kounkel et al. 2018) and OSCs (Soubiran
et al. 2018; Dias et al. 2018) close to the Sun have been
analyzed, and new OSCs have been detected (Beccari et al. 2018).
It is also interesting to note a recent review by Krumholz et al.
(2018), where the current status of research on the long-term
evolution of clusters is reflected.

The kinematics of OSCs with the proper motions and trigonometric
parallaxes calculated from Gaia DR2 data was analyzed in Bobylev
and Bajkova (2019), where the mean line-of-sight velocities of
clusters were taken mostly from the MWSC (Milky Way Star Clusters)
catalogue (Kharchenko et al. 2013). A catalogue by Soubiran et al.
(2018), where new mean values of the line-of-sight velocities were
calculated exclusively from the Gaia DR2 catalogue, has recently
been published. Using these data to determine the Galactic
rotation parameters and to solve other stellar-astronomy problems
is of great interest.

This paper is a continuation of the study by Bobylev and Bajkova
(2019) based on new mean values of the line-of-sight velocities
for OSCs, with great attention being given to studying the quality
of the new line-of-sight velocities for OSCs, the quality of the
distance scale, and the systematics of the Gaia DR2 catalogue.

 \section*{METHOD}
We know three stellar velocity components from observations: the
line-of-sight velocity $V_r$ and the two tangential velocity
components $V_l=4.74r\mu_l\cos b$ and $V_b=4.74r\mu_b$ along the
Galactic longitude $l$ and latitude $b,$ respectively, expressed
in km s$^{-1}$. Here, the coefficient 4.74 is the ratio of the
number of kilometers in an astronomical unit to the number of
seconds in a tropical year, and $r$ is the stellar heliocentric
distance in kpc. The proper motion components $\mu_l\cos b$ and
$\mu_b$ are expressed in mas yr$^{-1}$. To determine the
parameters of the Galactic rotation curve, we use the equations
derived from Bottlinger’s formulas, in which the angular velocity
$\Omega$ is expanded into a series to terms of the second order of
smallness in $r/R_0$:
\begin{equation}
 \begin{array}{lll}
 V_r=-U_\odot\cos b\cos l-V_\odot\cos b\sin l-W_\odot\sin b  \\
 +R_0(R-R_0)\sin l\cos b\Omega^\prime_0+0.5R_0(R-R_0)^2\sin l\cos b\Omega^{\prime\prime}_0,
 \label{EQ-1}
 \end{array}
 \end{equation}
 \begin{equation}
 \begin{array}{lll}
  V_l= U_\odot\sin l-V_\odot\cos l-r\Omega_0\cos b  \\
 +(R-R_0)(R_0\cos l-r\cos b)\Omega^\prime_0+0.5(R-R_0)^2(R_0\cos l-r\cos b)\Omega^{\prime\prime}_0 \\
 -r\cos l\sin b\omega_1-r\sin l\sin b\omega_2,
 \label{EQ-2}
 \end{array}
 \end{equation}
 \begin{equation}
 \begin{array}{lll}
 V_b=U_\odot\cos l\sin b + V_\odot\sin l \sin b-W_\odot\cos b\\
 -R_0(R-R_0)\sin l\sin b\Omega^\prime_0-0.5R_0(R-R_0)^2\sin l\sin b\Omega^{\prime\prime}_0\\
 +r\sin l\omega_1-r\cos l\omega_2,
 \label{EQ-3}
 \end{array}
 \end{equation}
where $R$ is the distance from the star to the Galactic rotation
axis:
  \begin{equation}
 R^2=r^2\cos^2 b-2R_0 r\cos b\cos l+R^2_0.
 \end{equation}
The quantity $\Omega_0$ is the angular velocity of Galactic
rotation at the solar distance $R_0,$ the parameters
$\Omega^{\prime}_0$ and $\Omega^{\prime\prime}_0$ are the
corresponding derivatives of this angular velocity, and the linear
rotation velocity of the Galaxy is $V_0=|R_0\Omega_0|.$

A periodicity with a wavelength of 2--3 kpc, which is attributable
to the influence of the Galactic spiral density wave, is known to
be observed (see, e.g., Rastorguev et al. 2017) on the Galactic
rotation curve whose parameters are determined from sufficiently
young disk objects. Therefore, in the first step it is convenient
to have a smooth Galactic rotation curve (which is provided by
expanding the angular velocity of rotation to the second order)
and then to study the influence of perturbations from the spiral
density wave separately. This approach was applied in Bobylev and
Bajkova (2018b, 2019).

Apart from the the rotation around the Galactic $z$ axis
(described by the parameter $\Omega$), here we also consider the
angular velocities of rotation around the $x$ and $y$ axes
described by the parameters $\omega_1$ and $\omega_2$,
respectively. Note that the parameters $\omega_1$ and $\omega_2$
can be found only from Eqs. (2) and (3). Young thin disk OSCs with
small values of sin b constitute the bulk of objects in our list.
The application of Eq. (3) is inefficient in searching for
$\Omega^{\prime}_0$ and $\Omega^{\prime\prime}_0$ based on young
objects. However, Eq. (3) is of great interest in searching for
the parameters $\omega_1$ and $\omega_2$, because here there is no
factor $\sin b$ at these two unknowns.

We known a number of present-day studies devoted to determining
the mean distance from the Sun to the Galactic center using its
individual determinations in the last decade by independent
methods. For example, $R_0=8\pm0.2$ kpc (Vall\'ee 2017),
$R_0=8.4\pm0.4$ kpc (de Grijs and Bono 2017), or $R_0=8.0\pm0.15$
kpc (Camarillo et al. 2018). Based on these reviews, here we
adopted $R_0=8\pm0.15$ kpc.

The kinematic parameters are determined by solving the conditional
equations (1)--(3) by the least-squares method (LSM). We use
weights of the form $w_r=S_0/\sqrt {S_0^2+\sigma^2_{V_r}}$ and
 $w_l=S_0/\sqrt {S_0^2+\sigma^2_{V_l}},$ where $S_0$ is the ``cosmic'' dispersion,
 $\sigma_{V_r}$ and $\sigma_{V_l}$ are the dispersions of the corresponding observed
velocities. $S_0$ is comparable to the root-mean-square residual
$\sigma_0$ (the error per unit weight) that is calculated when
solving the conditional equations (1)--(3). $S_0$ depends strongly
on the age of objects; in this paper we assign it to be close to
the value corresponding to the error per unit weight $\sigma_0$
found in advance. The LSM solutions were sought in two iterations
with the elimination of the residuals according to the $3\sigma$
criterion.

 \section*{DATA}
 \subsection*{Proper Motions and Line-of-Sight Velocities of OSCs}
The main source of the mean proper motions and parallaxes
calculated from Gaia DR2 data for us was the paper by
Cantat-Gaudin et al. (2018), where these quantities were
determined for 1229OSCs. The parameters of several more OSCs were
taken from Babusiaux et al. (2018), where they were calculated
only from Gaia DR2 data based on a large number of most probable
cluster members.

We took the mean heliocentric line-of-sight velocities of 953 OSCs
from the MWSC (Milky Way Star Clusters) catalogue (Kharchenko et
al. 2013) and, in several cases, from Kuhn et al. (2018),
Babusiaux et al. (2018), Casamiquela et al. (2016), Conrad et al.
(2014), and Mermilliod et al. (2008). In addition, we used the
line-of-sight velocities of 861 OSCs calculated by Soubiran et al.
(2018) exclusively from Gaia DR2 data.

Quite a few OSCs with measured line-of-sight velocities are common
to MWSC and Soubiran et al. (2018). As the comparison of these and
other catalogues made by Soubiran et al. (2018) showed, the
quality of the line-of-sight velocities for OSCs calculated from
Gaia DR2 data depends very strongly on the number of stars in each
cluster used for averaging. There is significant disagreement with
other measurements. For example, in some cases, the line-of-sight
velocity differences $|\Delta V_r|$ can exceed 50 km s$^{-1}$
(Fig. 5 in Soubiran et al. (2018)). Nevertheless, Soubiran et al.
(2018) identified $\sim$400 OSCs with a high quality of the
line-of-sight velocities (their mean error is 0.5 km s$^{-1}$)
calculated, on average, from seven stars.

Bobylev and Bajkova (2019) determined the Galactic rotation and
spiral density wave parameters based on a sample of young OSCs
from the catalogue by Cantat-Gaudin et al. (2018), where the
line-of-sight velocities exclusively from the MWSC catalogue was
used. In this paper we give preference to the line-of-sight
velocities from Soubiran et al. (2018). Thus, only if there is no
line-of-sight velocity in the list by Soubiran et al. (2018) do we
take its value from the MWSC catalogue.

The age estimates for the overwhelming majority of OSCs were taken
from the MWSC catalogue (Kharchenko et al. 2013). In isolated
cases, we invoked the estimates from the WEBDA electronic database
(https://www.univie.ac.at/webda/). Less than 100 OSCs have no age
estimates, 60 of them are recently discovered Gulliver clusters
(Cantat-Gaudin et al. 2018).

In this paper, in total, we consider OSCs with relative parallax
errors $\sigma_\pi/\pi<30\%$, where the dispersions $\sigma_\pi$
were taken from column 109 in the catalogue by Cantat-Gaudin et
al. (2018). There are a total of 1052 such OSCs of various ages
with measured proper motions and parallaxes and 863 of then also
have line-of-sight velocity estimates.

 \subsection*{Correction to the Gaia DR2 Parallaxes}
The presence of a systematic offset $\Delta\pi=-0.029$ mas in the
Gaia DR2 parallaxes with respect to an inertial reference frame
was first pointed out by Lindegren et al. (2018). Here the minus
means that the correction should be added to the Gaia DR2 stellar
parallaxes to reduce them to the standard. Arenou et al. (2018)
compared the Gaia DR2 parallaxes with 29 independent distance
scales that confirm the presence of an offset in the Gaia DR2
parallaxes $\Delta\pi\sim-0.03$ mas.

Stassun and Torres (2018) found the correction
$\Delta\pi=-0.082\pm0.033$ mas by comparing the parallaxes of 89
detached eclipsing binaries with their trigonometric parallaxes
from the Gaia DR2 catalogue.

By comparing the Gaia DR2 trigonometric parallaxes and photometric
parallaxes of 94 OSCs, Yalyalieva et al. (2018) found the
correction $\Delta\pi=-0.045\pm0.009$ mas. The high accuracy of
this estimate is attributable to the high accuracy of the
photometric distance estimates for OSCs obtained by invoking
first-class infrared photometric surveys, such as IPHAS, 2MASS,
WISE, and Pan-STARRS.

Riess et al. (2018) obtained an estimate of
$\Delta\pi=-0.046\pm0.013$ mas based on a sample of 50 long-period
Cepheids when comparing their parallaxes with those from the Gaia
DR2 catalogue. The photometric parameters of these Cepheids
measured from the Hubble Space Telescope were used.

By comparing the distances of $\sim$3000 stars from the APOKAS-2
catalogue (Pinsonneault et al. 2018) belonging to the red giant
branch with the Gaia DR2 data, Zinn et al. (2018) found the
correction $\Delta\pi=-0.053\pm0.003$ mas. These authors also
obtained a close value $\Delta\pi=-0.050\pm0.004$ mas, by
analyzing stars belonging to the red giant clump. The distances to
such stars were estimated from asteroseismic data. According to
these authors, the parallax errors here are approximately equal to
the errors in estimating the stellar radius and are, on average,
1.5\%. Such small errors in combination with the enormous number
of stars allowed $\Delta\pi$ to be determined with a high
accuracy.

The listed results lead to the conclusion that the trigonometric
parallaxes of stars from the Gaia DR2 catalogue should be
corrected by applying a small correction. We will be oriented to
the results of Yalyalieva et al. (2018), Riess et al. (2018), and
Zinn et al. (2018), which look most reliable. Thus,we apply a
correction of 0.050 mas to all the original parallaxes of OSCs.

 \begin{table}[t]
 \caption[]{\small
The Galactic rotation parameters found only from the line-of-sight
velocities $V_r$ (Eq. (1)) based on OSCs of various ages with
relative trigonometric parallax errors less than 15\%
 }
  \begin{center}  \label{t:00}
  \small
  \begin{tabular}{|l|r|r|r|r|r|}\hline
    Parameters                   &       All ages & $0<\log t\leq8$ & $8<\log t\leq9$ &      $9<\log t$ \\\hline
    $U_\odot,$    km s$^{-1}$    &  $11.4\pm0.8$  &  $10.7\pm1.4$  &  $11.7\pm1.1$  & $ 11.9\pm3.3$  \\
    $V_\odot,$    km s$^{-1}$    &  $13.1\pm0.8$  &  $13.4\pm1.5$  &  $13.0\pm1.2$  & $ 13.1\pm3.2$  \\
  $\Omega^{'}_0,$ km s$^{-1}$ kpc$^{-2}$ & $-3.97\pm0.10$ & $-3.85\pm0.19$ & $-4.05\pm0.14$ & $-3.85\pm0.38$ \\
 $\Omega^{''}_0,$ km s$^{-1}$ kpc$^{-3}$ & $0.45\pm0.088$ & $ 0.49\pm0.23$ & $ 0.57\pm0.13$ & $ 0.27\pm0.21$ \\
   $\sigma_0,$    km s$^{-1}$    &           15.4 &           14.2 &           13.5 &           20.4 \\
     $N_\star$                   &            856 &            273 &            434 &            110 \\
  \hline
 \end{tabular}\end{center}
  {\small $N_\star$ is the number of clusters used.}
  \end{table}
 \begin{table}[t]
 \caption[]{\small
The Galactic rotation parameters found from OSCs of various ages
with relative trigonometric parallax errors less than 15\%
 }
  \begin{center}  \label{t:01}
  \small
  \begin{tabular}{|l|r|r|r|r|r|}\hline
    Parameters                   &         All ages & $0<\log t\leq8$ &  $8<\log t\leq9$ &     $9<\log t$ \\\hline
    $U_\odot,$    km s$^{-1}$    &    $11.7\pm1.0$  &  $ 9.0\pm1.4$  &   $10.2\pm1.3$  & $ 15.1\pm4.3$  \\
    $V_\odot,$    km s$^{-1}$    &    $13.1\pm1.0$  &  $13.9\pm1.4$  &   $13.4\pm1.5$  & $ 13.6\pm3.9$  \\
  $\Omega^{'}_0,$ km s$^{-1}$ kpc$^{-2}$ &   $-3.99\pm0.14$ & $-4.05\pm0.23$ & $-4.21\pm0.19$  & $-3.87\pm0.46$ \\
 $\Omega^{''}_0,$ km s$^{-1}$ kpc$^{-3}$ &    $0.37\pm0.10$ & $ 1.18\pm0.37$ & $ 0.67\pm0.21$  & $ 0.21\pm0.23$ \\
   $\sigma_0,$    km s$^{-1}$    &             13.5 &            9.4 &            13.5 &           20.3 \\
     $N_\star$                   &              456 &            127 &             238 &             74 \\
 \hline
    $U_\odot,$    km s$^{-1}$    &   $ 8.60\pm0.54$ & $ 7.68\pm0.63$   & $  9.02\pm0.81$  & $ 11.7\pm2.4$  \\
    $V_\odot,$    km s$^{-1}$    &   $ 9.36\pm0.70$ & $ 8.15\pm0.85$   & $ 10.59\pm1.03$  & $ 10.1\pm3.5$  \\
  $\Omega_0,$     km s$^{-1}$ kpc$^{-1}$ & $28.78\pm0.30$   & $29.70\pm0.36$   & $ 28.64\pm0.45$  & $ 29.2\pm1.2$  \\
  $\Omega^{'}_0,$ km s$^{-1}$ kpc$^{-2}$ & $-3.988\pm0.081$ & $-4.061\pm0.036$ & $-4.107\pm0.123$ & $-3.67\pm0.34$ \\
 $\Omega^{''}_0,$ km s$^{-1}$ kpc$^{-3}$ & $ 0.637\pm0.039$ & $ 0.677\pm0.032$ & $ 0.713\pm0.063$ & $ 0.40\pm0.12$ \\
   $\sigma_0,$    km s$^{-1}$    &           12.72  &             8.5  &           13.0   &           18.8 \\
     $N_\star$                   &            1052  &             345  &            518   &            122 \\
      $z_\odot,$  pc             &       $-10\pm5$  &       $-21\pm4$  &       $-14\pm5$  &     $ 32\pm30$ \\
 $(\Omega^{'}_0)_{V_r}/(\Omega^{'}_0)_{V_l}$ & $1.00\pm0.04$ &  $1.00\pm0.06$ & $1.03\pm0.05$ &  $1.05\pm0.16$  \\
  \hline
 \end{tabular}\end{center}

 {\small
 The results obtained only from the line-of-sight velocities $V_r$
(Eq. (1)) and only from the component $V_l$ (Eq. (2)) are given in
the upper and lower parts, respectively, $N_\star$ is the number
of clusters used.}
 \end{table}

 \section*{RESULTS}
Table 1 gives the kinematic parameters derived by solving Eqs. (1)
for various age groups. Note that the velocity $W_\odot$ is
determined very poorly in this case of using only the
line-of-sight velocities and, therefore, we eliminate it as a
parameter being determined by taking it to be 7 km s$^{-1}$. OSCs
with relative trigonometric parallax errors less than 15\% were
used here. We used all of the OSCs with a nonzero number of
probable cluster members from which the mean line-of-sight
velocities were calculated. In the case of nonzero line-of-sight
velocities in two catalogues, MWSC and Soubiran et al. (2018),
preference was given to the latter.

The error per unit weight $\sigma_0$ that we find when solving the
conditional equations (1)--(3) characterizes the residual velocity
dispersion for OSCs averaged over three directions. The residual
velocity dispersion for hydrogen clouds in the Galactic disk is
known to be $\sim$5 km s$^{-1}$. The residual velocity dispersion
for OB stars lies in the range 8--10 km s$^{-1}$. One might expect
the velocity dispersion of relatively young OSCs, $0<\log t\leq
8$, to be 8--10 km s$^{-1}$. However, for this sample of OSCs this
quantity turned out to be considerably larger than the expected
one, $\sigma_0=14.2$ km s$^{-1}$. To improve the quality of OSC
samples with measured line-of-sight velocities, below we proceed
as did Soubiran et al. (2018). More specifically, we use OSCs with
the mean line-of-sight velocities calculated from at least three
probable cluster members.

Table 2 gives the kinematic parameters derived by separately
solving Eqs. (1) and (2) for various age groups. We used OSCs with
relative trigonometric parallax errors less than 15\%. When
seeking a solution only from the line-of-sight velocities, we used
OSCs with the mean values calculated from at least three probable
cluster members. As can be seen from Table 2, with this approach
the values of $\sigma_0$ determined from various types of
velocities came closer together, although the values derived from
the line-of-sight velocities slightly exceed those found from the
proper motions.

As an additional characteristic of the spatial distribution of the
sample Table 2 gives the mean $z$ coordinate that is designated as
$z_\odot$ and reflects the well-known fact of the Sun’s elevation
above the Galactic plane. A review of its determinations can be
found, for example, in Bobylev and Bajkova (2016), where
$z_\odot=-16\pm2$ pc was found from several samples of young
objects. As follows from Table 2, the values of $z_\odot$ found
from OSCs with ages $\log t<9.0$ are in good agreement with the
known results. Only $z_\odot$ found from the sample of oldest OSCs
agrees poorly with other determinations. Furthermore, the OSCs
with a logarithm of the age more than nine have a very large error
per unit weight, i.e., a large velocity dispersion; therefore, we
do not use these OSCs below. An analysis of Tables 1 and 2 (mostly
$\sigma_0$) also leads us to conclude that the Galactic rotation
parameters are determined more accurately using only their proper
motions than only their line-of-sight velocities.

The last row in Table 2 gives the ratio of the first derivative of
the angular velocity found using only the line-of-sight
velocities, $(\Omega^{'}_0)_{V_r}$, to that found using only the
proper motions, $(\Omega^{'}_0)_{V_l}.$ This method is based on
the fact that the errors in the line-of-sight velocities do not
depend on the distance errors, while the errors in the tangential
components depend on them. Therefore, comparing the values of
$\Omega_0'$ found by various methods allows the correction factor
of the distance scale $p$ to be found (Zabolotskikh et al. 2002;
Rastorguev et al. 2017; Bobylev and Bajkova 2017a); in our case,
$p=(\Omega^{'}_0)_{V_r}/(\Omega^{'}_0)_{V_l}$. The error in $p$
was calculated based on the relation
$\sigma^2_p=(\sigma_{\Omega'_{0V_r}}/\Omega'_{0V_l})^2+
     (\Omega'_{0V_r}\cdot\sigma_{\Omega'_{0V_l}}/\Omega'^2_{0V_l})^2.$
Having analyzed more than 50 000 stars from the TGAS catalogue
(Brown et al. 2016), Bobylev and Bajkova (2017a) obtained an
estimate of $p=0.97\pm0.04$ by this method. According to the
results from Table 2, we can unambiguously conclude that the
distance scale factor is equal to unity. Therefore, the distances
used do not need any correction factor.

\begin{figure}[p]
{\begin{center}
   \includegraphics[width=0.8\textwidth]{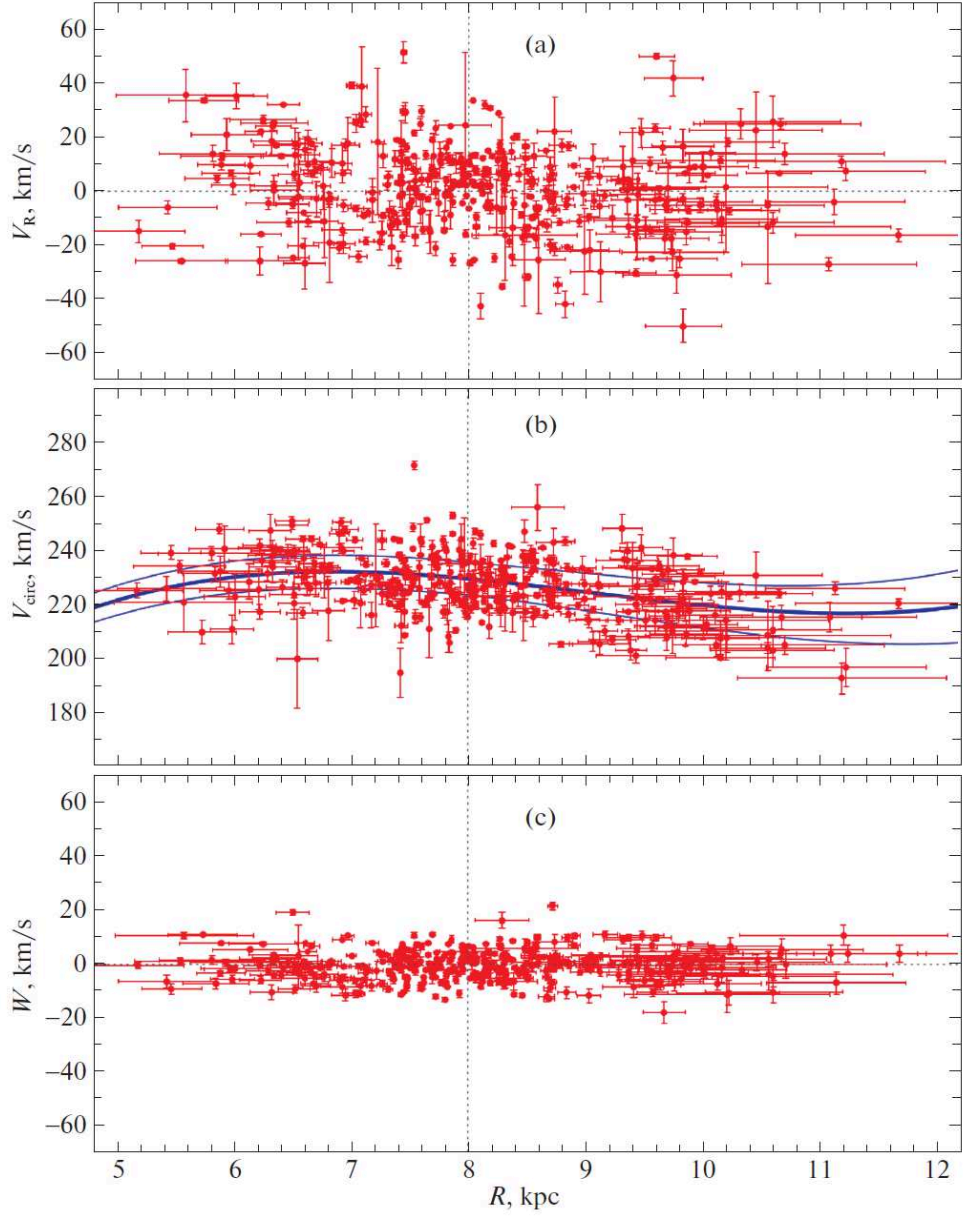}
 \caption{
Radial (a), tangential (b), and vertical (c) velocities for the
sample of 384 OSCs with measured space velocities versus
Galactocentric distance; the vertical dotted line marks the Sun’s
position.
  } \label{f1}
\end{center}}
\end{figure}

We then selected OSCs with the constraints on the age $\log t<9$
and the relative parallax error $\sigma_\pi/\pi<30\%.$ There were
a total of 930 such clusters. Among them there are 384 OSCs with
the mean line-of-sight velocities that were calculated using at
least three probable cluster members. Based on this sample of
OSCs, we found the following kinematic parameters by
simultaneously solving all Eqs.(1)--(3):
 \begin{equation}
 \label{solution-best}
 \begin{array}{lll}
 (U_\odot,V_\odot,W_\odot)=(8.53,11.22,7.83)\pm(0.38,0.46,0.32)~\hbox{km s$^{-1}$},\\
      \Omega_0 =~28.71\pm0.22~\hbox{km s$^{-1}$ kpc$^{-1}$},\\
  \Omega^{'}_0 =-4.100\pm0.058~\hbox{km s$^{-1}$ kpc$^{-2}$},\\
 \Omega^{''}_0 =~0.736\pm0.033~\hbox{km s$^{-1}$ kpc$^{-3}$},
 \end{array}
 \end{equation}
where the error per unit weight is $\sigma_0=9.71$~km s$^{-1}$,
the linear rotation velocity of the Galaxy is $V_0=229.7\pm4.6$ km
s$^{-1}$ at the solar distance, and the Oort constants are $A=
16.40\pm0.23$ km s$^{-1}$ kpc$^{-1}$ and $B=-12.31\pm0.32$ km
s$^{-1}$ kpc$^{-1}$.

For 384 OSCs with measured space velocities (these OSCs were used
in seeking the solution (5)) their Galactocentric radial, $V_R,$
tangential, $V_{circ},$ and vertical, $W,$ velocities are plotted
against the distance $R$ in Fig.~1. We see that the greatest and
smallest velocity dispersions are observed in the radial (Galactic
center–anticenter) and vertical directions, respectively. The
velocity dispersions calculated from this sample of OSCs are:
 $(\sigma_{V_R},\sigma_{\Delta V_{circ}},\sigma_W)=(15.8,11.3,5.5)$
km s$^{-1}$.

This sample contains many relatively young OSCs and, therefore,
low-amplitude waves with a wavelength of 2--3 kpc due to the
influence of the Galactic spiral density wave are visible on all
three graphs. Such periodicities in the velocities of OSCs have
recently been studied in more detail by Bobylev and Bajkova
(2019).

Based on the same sample of 930 OSCs with $\lg t<9.0$ and
$\sigma_\pi/\pi<30\%$, we found the following kinematic parameters
using only two equations, (2) and (3):
 \begin{equation}
 \label{solution-w1213}
 \begin{array}{lll}
 (U_\odot,V_\odot,W_\odot)= (7.90,9.61,7.79)\pm(0.40,0.53,0.29)~\hbox{km s$^{-1}$},\\
      \Omega_0 =~28.88\pm0.22~\hbox{km s$^{-1}$ kpc$^{-1}$},\\
  \Omega^{'}_0 =-4.078\pm0.061~\hbox{km s$^{-1}$ kpc$^{-2}$},\\
 \Omega^{''}_0 =~0.684\pm0.033~\hbox{km s$^{-1}$ kpc$^{-3}$},\\
      \omega_1 = 0.48\pm0.15~\hbox{km s$^{-1}$ kpc$^{-1}$},\\
      \omega_2 = 0.32\pm0.20~\hbox{km s$^{-1}$ kpc$^{-1}$},
 \end{array}
 \end{equation}
where the error per unit weight is $\sigma_0=8.5$ km s$^{-1}$.
Interestingly, $\omega_1$ here differs significantly from zero.

 \section*{DISCUSSION}
 \subsection*{Rotation around the $z$ Axis}
Based on 130 masers with measured VLBI trigonometric parallaxes,
Rastorguev et al. (2017) found the solar velocity components
$(U_\odot,V_\odot)=(11.40,17.23)\pm(1.33,1.09)$ km s$^{-1}$ and
the following parameters of the Galactic rotation curve:
 $\Omega_0=28.93\pm0.53$~km s$^{-1}$ kpc$^{-1}$,
 $\Omega^{'}_0=-3.96\pm0.07$~km s$^{-1}$ kpc$^{-2}$ and
 $\Omega^{''}_0=0.87\pm0.03$~km s$^{-1}$ kpc$^{-3},$ where
 $V_0=243\pm10$~km s$^{-1}$ (for $R_0=8.40\pm0.12$~kpc found).

Based on a sample of 495 OB stars with proper motions from the
Gaia DR2 catalogue, Bobylev and Bajkova (2018b) found the
following kinematic parameters:
$(U,V,W)_\odot=(8.16,11.19,8.55)\pm(0.48,0.56,0.48)$~km s$^{-1}$,
      $\Omega_0=28.92\pm0.39$~km s$^{-1}$ kpc$^{-1}$,
  $\Omega^{'}_0=-4.087\pm0.083$~km s$^{-1}$ kpc$^{-2}$ and
 $\Omega^{''}_0=0.703\pm0.067$~km s$^{-1}$ kpc$^{-3}$, where
$V_0=231\pm5$~km s$^{-1}$ (for the adopted $R_0=8.0\pm0.15$ kpc).

Based on a sample of 326 young OSCs with proper motions from the
Gaia DR2 catalogue and line-of-sight velocities from the MWSC
catalogue, Bobylev and Bajkova (2019) found the following
kinematic parameters:
$(U,V,W)_\odot=(7.88,11.17,8.28)\pm(0.48,0.63,0.45)$~km s$^{-1}$,
 $\Omega_0 =29.34\pm0.31$~km s$^{-1}$ kpc$^{-1}$,
 $\Omega^{'}_0=-4.012\pm0.074$~km s$^{-1}$ kpc$^{-2}$ and
 $\Omega^{''}_0=0.779\pm0.062$~km s$^{-1}$ kpc$^{-3}$, where
$V_0=235\pm5$~km s$^{-1}$ ($R_0=8.0\pm0.15$~kpc).

Note that the kinematic parameters corresponding to our solutions
(5) and (6) are in best agreement with these three results.
Moreover, the errors in the parameters being determined are
smallest in the solution (5). Thus, we conclude that the best
solution for the sought-for kinematic parameters is the solution
(5) obtained from the sample of 930 OSCs satisfying the
constraints on the age $\lg t<9.0$ and the relative trigonometric
parallax error $<15\%,$ which also includes the line-of-sight
velocities of 384 OSCs calculated from at least three probable
members of the corresponding clusters.

\subsection*{Rotation around the $x$ and $y$ Axes}
Liu et al. (2017) performed a kinematic analysis of $\sim$23 000
K--M giants from the TGAS catalogue (Brown et al. 2016) based on
the Ogorodnikov--Milne model and found a nonzero component
$\omega_2=-0.38\pm0.15$ mas yr$^{-1}$ (rotation around the
Galactic y axis). They interpreted this as a possible residual
rotation in the TGAS frame or the presence of problems in the
kinematic model.

Our analysis of the kinematics of OSCs does not confirm such a
rapid rotation. Indeed, the mean distance for the sample of stars
used in obtaining the solution (6) is 2.0 kpc; then
$\omega_1=0.051\pm0.016$ mas yr$^{-1}$ and
$\omega_2=0.034\pm0.021$ mas yr$^{-1}$. We see that here
$\omega_2$ differs from the result of Liu et al. (2017) by an
order of magnitude.

As Lindegren et al. (2018) showed, the Gaia DR2 frame has no
rotation relative to the system of quasars within 0.15 mas
yr$^{-1}$, with the greatest effect manifesting itself in the
region of bright $(G<12^m)$ stars. The mean proper motions of the
OSCs under consideration seem to have been calculated from a
considerable number of precisely bright stars. Thus, on the one
hand, the value of $\omega_1=0.051\pm0.016$ mas yr$^{-1}$ found in
this paper may be a consequence of a slight residual rotation of
the Gaia DR2 frame relative to the extragalactic reference frame.
On the other hand, the slight rotation around the Galactic $x$
axis with an angular velocity of $0.48\pm0.15$ km s$^{-1}$
kpc$^{-1}$ found in this paper may somehow be related to the
presumed precession/rotation of the warped Galactic disk. Various
authors have attempted to detect this kinematic effect in the
kinematics of stars and clusters over many years. However, at
present there is no agreement between the results obtained. For
example, using the proper motions of O--B5 stars, Miyamoto and Zhu
(1998) found rotation of this system of stars around the Galactic
$x$ axis with an angular velocity of about 4 km s$^{-1}$
kpc$^{-1}$ based on the simplest solid-body rotation model. Based
on the proper motions of $\sim$80 000 red giant clump stars,
Bobylev (2010) found rotation of this system of stars around the x
axis with an angular velocity of about $-4$ km s$^{-1}$
kpc$^{-1}$, while based on a sample of classical Cepheids, Bobylev
(2013) found rotation around the x axis with an angular velocity
of $-15\pm5$ km s$^{-1}$ kpc$^{-1}$.

On the whole, various observations confirm the asymmetry in the
vertical velocities of stars (L\'opez-Corredoira et al. 2014;
Romero-G\'omez et al. 2018), but the application of a complex disk
precession model is required to describe the phenomenon. Based on
a simplified approach, L\'opez-Corredoira et al. (2014) obtained
an estimate of the rotation around the $x$ axis with an angular
velocity of $\sim-2$ km s$^{-1}$ kpc$^{-1}$. While analyzing the
Gaia DR2 data, Romero-G\'omez et al. (2018) showed that the
observed structure of the vertical velocities is complex and
depends strongly on the sample age.

 \section*{CONCLUSIONS}
Thus, based on published data, we selected a sample of more than
1000 OSCs with their proper motions and parallaxes from the Gaia
DR2 catalogue and their line-of-sight velocities from the Gaia DR2
(predominantly) and MWSC catalogues. Following the latest results
of an analysis of the zero point for the Gaia DR2 parallax scale,
we calculated the distances to OSCs by adding the correction
$\Delta\pi=0.050$ mas to the original mean values of their
parallaxes.

We studied the joint and separate solutions of the basic kinematic
equations when determining the Galactic rotation parameters. We
showed that the Galactic rotation parameters are determined more
accurately from OSCs located within 4--5 kpc of the Sun using only
their proper motions than only their line-of-sight velocities. We
conclude that the distance scale factor is virtually equal to
unity, $p=1.00\pm0.04.$ Therefore, the distances calculated using
the parallaxes from the Gaia DR2 catalogue do not need any
correction factor.

Nevertheless, the best (with the smallest error per unit weight
and with the smallest errors of the parameters being determined)
solution was obtained as a result of the simultaneous solution
based on a sample of 930 OSCs selected under the constraint on the
age $\log t<9.0$ with relative trigonometric parallax errors less
than 30\%. In this solution we used 384 OSCs with the mean
line-of-sight velocities calculated from at least three probable
cluster members. The parameters found are reflected in the
solution (5).

Having analyzed the proper motions of 930 OSCs, we established
that, apart from the rotation around the Galactic $z$ axis (the
well-known Galactic rotation), there is rotation of the entire
sample around the Galactic $x$ axis with an angular velocity
$\omega_1=0.48\pm0.15$ km s$^{-1}$ kpc$^{-1}$ differing
significantly from zero. This quantity can also be expressed in
angular units, given the mean distance of the OSC sample; then
$\omega_1=0.051\pm0.016$ mas yr$^{-1}$. This rotation can be both
a kinematic peculiarity of OSCs and a consequence of the slight
residual rotation of the Gaia DR2 frame relative to the
extragalactic reference frame. Of course, this effect should be
studied and confirmed on greater statistics.

 \section*{ACKNOWLEDGMENTS}
We are grateful to the referees for their useful remarks that
contributed to an improvement of the paper.

 \section*{FUNDING}
This work was supported in part by Basic Research Program P-28 of
the Presidium of the Russian Academy of Sciences, the subprogram
``Cosmos: Studies of Fundamental Processes and their
Interrelations''.

 \bigskip \bigskip\medskip{\bf REFERENCES}{\small

1. F. Arenou, X. Luri, C. Babusiaux, C. Fabricius, A. Helmi, T.
Muraveva, A. C. Robin, F. Spoto, et al. (Gaia Collab.), Astron.
Astrophys. 616, 17 (2018).

2. C. Babusiaux, F. van Leeuwen, M. A. Barstow, C. Jordi, A.
Vallenari, A. Bossini, A. Bressan, T. Cantat-T. Gaudin, et al.
(Gaia Collab.), Astron. Astrophys. 616, 10 (2018).

3. G. Beccari, H. M. J. Boffin, T. Jerabkova, N. J. Wright, V. M.
Kalari, G. Carraro, G. de Marchi, and W.-J. de Wit, Mon. Not. R.
Astron. Soc. 481, L11 (2018).

4. V. V. Bobylev, Astron. Lett. 36, 634 (2010).

5. V. V. Bobylev, Astron. Lett. 39, 819 (2013).

6. V. V. Bobylev and A. T. Bajkova, Astron. Lett. 42, 1 (2016).

7. V. V. Bobylev and A. T. Bajkova, Astron. Lett. 44, 184 (2018a).

8. V. V. Bobylev and A. T. Bajkova, Astron. Lett. 44, 675 (2018b).

9. V. V. Bobylev and A. T. Bajkova, Astron. Lett. 45, 109 (2019).

10. A. G. A. Brown, A. Vallenari, T. Prusti, J. de Bruijne, F.
Mignard, R. Drimmel, C. Babusiaux, C. A. L. Bailer-Jones, et al.
(Gaia Collab.), Astron. Astrophys. 595, 2 (2016).

11. A. G. A. Brown, A. Vallenari, T. Prusti, de Bruijne, C.
Babusiaux, C. A. L. Bailer-Jones, M. Biermann, D.W. Evans, et al.
(Gaia Collab.), Astron. Astrophys. 616, 1 (2018).

12. T. Camarillo, M. Varun, M. Tyler, and R. Bharat, Publ. Astron.
Soc. Pacif. 130, 4101 (2018).

13. T. Cantat-Gaudin, C. Jordi, A. Vallenari, A. Bragaglia, L.
Balaguer-Nunez, C. Soubiran, et al., Astron. Astrophys. 618, A93
(2018).

14. L. Casamiquela, R. Carrera, C. Jordi, L. Balaguer-Nunez, E.
Pancino, S. L. Hidalgo, C. E. Martinez-V\'azquez, S. Murabito, et
al., Mon. Not. R. Astron. Soc. 458, 3150 (2016).

15. C. Conrad, R.-D. Scholz, N. V. Kharchenko, A. E. Piskunov, E.
Schilbach, S. R\"oser, C. Boeche, G. Kordopatis, et al., Astron.
Astrophys. 562, 54 (2014).

16. W. S. Dias, H. Monteiro, J. R. D. L\'epine, R. Prates, C. D.
Gneiding, and M. Sacchi, Mon. Not. R. Astron. Soc. 481, 3887
(2018).

17. E. Franciosini, G. G. Sacco, R. D. Jeffries, F. Damiani, V.
Roccatagliata, D. Fedele, and S. Randich, Astron. Astrophys. 616,
12 (2018).

18. R. de Grijs and G. Bono, Astrophys. J. Suppl. Ser. 232, 22
(2017).

19. N. V. Kharchenko, A. E. Piskunov, E. Schilbach, S. R\"oser,
and R.-D. Scholz, Astron. Astrophys. 558, 53 (2013).

20. M. Kounkel, K. Covey, G. Su\'arez, C. Roman-Zuniga, J.
Hernandez, K. Stassun, K. O. Jaehnig, E. D. Feigelson, et al.,
Astron. J. 156, 84 (2018).

21. M. R. Krumholz, C. F. McKee, and J. Bland- Hawthorn, arXiv:
1812.01615 (2018).

22. M. A. Kuhn, L. A. Hillenbrand, A. Sills, E. D. Feigelson, and
K. V. Getman,Astrophys. J. 870, 32 (2018).

23. L. Lindegren, J. Hernandez, A. Bombrun, S. Klioner, U.
Bastian, M. Ramos-Lerate, A. de Torres, H. Steidelmuller, et al.
(Gaia Collab.), Astron. Astrophys. 616, 2 (2018).

24. N. Liu, Z. Zhu, J.-C. Liu, and C.-Y. Ding, Astron. Astrophys.
599, 140 (2017).

25. M. L\'opez-Corredoira, H. Abedi, F. Garz\'on, and F. Figueras,
Astron. Astrophys. 572, 101 (2014).

26. J. C. Mermilliod, M. Mayor, and S. Udry, Astron. Astrophys.
485, 303 (2008).

27. M. Miyamoto and Z. Zhu, Astron. J. 115, 1483 (1998).

28. M. H. Pinsonneault, Y. P. Elsworth, J. Tayar, A. Serenelli, D.
Stello, J. Zinn, S. Mathur, R. Garcia, et al., Astrophys. J.
Suppl. Ser. 239, 32 (2018).

29. A. S. Rastorguev, M. V. Zabolotskikh, A. K. Dambis, N. D.
Utkin, V. V. Bobylev, and A. T. Bajkova, Astrophys. Bull. 72, 122
(2017).

30. A. G. Riess, S. Casertano,W. Yuan, L. Macri, B. Bucciarelli,
M. G. Lattanzi, J.W. MacKenty, J. B. Bowers, et al., Astrophys. J.
861, 126 (2018).

31. V. Roccatagliata, G. G. Sacco, E. Franciosini, and S. Randich,
Astron. Astrophys. 617, L4 (2018).

32. M. Romero-G\'omez, C. Mateu, L. Aguilar, F. Figueras, and A.
Castro-Ginard, arXiv: 1812.07576 (2018).

33. C. Soubiran, T. Cantat-Gaudin, M. Romero-Gomez, L.
Casamiquela, C. Jordi, A. Vallenari, T. Antoja, L. Balaguer-Nunez,
et al., Astron. Astrophys. 619, A155 (2018).

34. K. G. Stassun and G. Torres, Astrophys. J. 862, 61 (2018).

35. J. P. Vall\'ee, Astrophys. Space Sci. 362, 79 (2017).

36. L. N. Yalyalieva, A. A. Chemel, E. V. Glushkova, A. K. Dambis,
and A. D. Klinichev, Astrophys. Bull. 73, 335 (2018).

37. M. V. Zabolotskikh, A. S. Rastorguev, and A. K. Dambis,
Astron. Lett. 28, 454 (2002).

38. E. Zari, H. Hashemi, A.G.A. Brown, K. Jardine, and P. T. de
Zeeuw, Astron. Astrophys. 620, A172 (2018).

39. J. C. Zinn, M. H. Pinsonneault, D. Huber, and D. Stello,
arXiv: 1805.02650 (2018).
  }
  \end{document}